\documentclass[prd,twocolumn,aps,preprintnumbers,letterpaper,floatfix]{revtex4}
\usepackage{amsmath, gensymb}
\usepackage{tipa}
\usepackage{amssymb}
\usepackage{graphicx}
\usepackage{bm}
\usepackage{enumerate}
\usepackage{subfigure}
\usepackage{color}
\newcommand{\ddd}{\displaystyle}

\newcommand{\be}{\begin{equation}}
\newcommand{\ee}{\end{equation}}
\newcommand{\bea}{\begin{eqnarray}}
\newcommand{\eea}{\end{eqnarray}}
\newcommand{\ba}{\begin{eqnarray}}
\newcommand{\ea}{\end{eqnarray}}

\renewcommand{\Im}{\mathrm{Im}\,}

\newcommand{\de}[2]{\delta_{#1 #2}}
\newcommand{\kh}{\hat k}

\newcommand{\GeV}{\mathrm{GeV}}
\newcommand{\MeV}{\mathrm{MeV}}

\begin{document}

\title{ Gravity waves generated by sounds from big bang phase transitions
}

\author{Tigran Kalaydzhyan and Edward Shuryak}
\affiliation{Department of Physics and Astronomy, Stony Brook University,\\ Stony Brook, New York 11794-3800, USA}

\date{\today}

\begin{abstract}
Inhomogeneities associated with the cosmological QCD and electroweak phase transitions produce hydrodynamical
perturbations, longitudinal sounds and rotations. It has been demonstrated by  Hindmarsh \textit{et al.} \cite{Hindmarsh:2013xza} that the sounds produce gravity waves (GW) well after the phase transition is over. We further argue
that, under certain conditions,
an  {\em inverse acoustic  cascade} may occur and move sound perturbations
  from the (UV) momentum scale at which the sound is originally produced
 to much smaller (IR) momenta. The weak turbulence regime of this cascade is studied via
  the Boltzmann equation, possessing stationary power and  time-dependent self-similar solutions.
  We suggest certain indices for the strong turbulence regime as well, into which the cascade eventually proceeds.
  Finally, we point out that two on shell sound waves can produce one on-shell gravity wave,  and we evaluate the rate
  of the process using a standard {\em sound loop diagram}.
 \end{abstract}

\maketitle

   \section{Introduction}

We think that our Universe was ``boiling" at its early stages at least three times: at
the initial  equilibration, when entropy was produced, and at electroweak and QCD phase transitions.
On general grounds, this boiling should have produced certain out-of-equilibrium effects.
It remains a great challenge for one to find a way to observe the consequences experimentally, or at least evaluate the magnitude of possible effects.

Thirty years ago, in a very influential paper, Witten \cite{Witten:1984rs}
 discussed the bubble dynamics, assuming that the cosmic QCD phase transition is of the first order. Among other things, he
  pointed out that  bubble coalescence or collisions  produce inhomogeneities of the energy density,
which  lead to the  gravity waves (GW) production.
These ideas were further developed by Hogan \cite{Hogan:1986qda},
who identified relevant frequencies and provided the first estimates
of the radiation intensity.

Hogan was also the first to  mention the subject of this work -- generation of the GW from the sound. Unfortunately, this idea was dormant for a very long time an was recently revived by  Hindmarsh \textit{et al.} \cite{Hindmarsh:2013xza}, who found the hydrodynamic sound waves to be the dominant  source of the GW (see also the later work \cite{Giblin:2014qia}).
This  paper triggered our interest in the subject.
 Hindmarsh \textit{et al.}, however, performed
  numerical simulations of (variant of) the electroweak (EW) phase transition, in the
  traditional first order transition setting.
  Thus, it is clear that previous calculations of the GW yield -- such as, e.g., Ref.~\cite{Caprini:2010xv} for the QCD transition --
  need to be strongly modified, including the dynamics of the sound waves.
    We return  to the discussion of Ref.~\cite{Hindmarsh:2013xza}  in Sec.~\ref{sec_Hindmarsh}.

  Our paper refers to both QCD and EW transitions,
  with emphasis on the former case, because of favorable observational prospects and our background.
The main point of our paper is that, given a huge dynamical range of the problem, it is clearly
impossible to cover  it  in a single numerical setting.
We suggest splitting the problem into distinct stages,
 each with its own physics, scales and technique. We list them starting from the  UV end of the spectrum, with momenta
 of the order of ambient temperature $k\sim T_c$,
 and ending at the IR end of the spectrum, $k\sim 1/t_{life}$, limited by the cosmological horizon (inverse to the Universe lifetime)
  at the radiation-dominated era:
 \begin{itemize}
 \item[(i)] production of sounds from inhomogeneities,\vspace{-0.2cm}
 \item[(ii)] inverse acoustic cascade, focusing sound-wave population
 toward small momenta, \vspace{-0.2cm}
 \item[(iii)] the final conversion of sounds into the GW.
 \end{itemize}

Stage (i) remains highly nontrivial, associated with the dynamical details of the
 QCD and EW phase transitions. We will not be able to provide definite predictions at this point;
 and only  make some comments on the current status
of the problem in Sec.~\ref{sec_qcd_1storder}.

 Stage (ii) will be our main focus. It is, in fact, amenable to perturbative studies
 of the acoustic inverse cascade, consisting of sound decay or scattering
 events. Those events are governed by the Boltzmann equation which has already been
  studied in the literature on acoustic turbulence, to certain extent.  The stationary attractor solutions -- known as Kolmogorov-Zakharov
  spectra -- can be identified, as can some time-dependent self-similar solutions describing a spectrum profile 
  moving across the dynamical range.
 Application of this theory allows us to see how small-amplitude sounds at the UV
 get self-focused at small $k$, tremendously
   amplifying the momentum density $n_k$ there.

 The  final step, (iii), can be treated directly via a standard on-shell process for the
 $sound+sound \rightarrow GW$ transition, to be calculated
   in Sec.~\ref{soundsound_toGW} via a sound loop diagram.
   Since it is proportional to squared density $(n_k)^2$, it can be
   amplified by an inverse acoustic cascade by a huge factor.

 Let us note that the studies of the QCD phase transition region,
    from the confined (or hadronic) phase to the deconfined quark-gluon plasma (QGP) phase, now constitute
     the mainstream of the heavy-ion physics.  Experiments, done mostly at the RHIC in Brookhaven and
     now at CERN LHC, revealed that the matter
   above and near the phase transition seems to be a nearly perfect liquid with a small viscosity.
A hydrodynamic description of the subsequent explosion -- sometimes called the little bang -- turns out to be very accurate.

   Furthermore, initial state fluctuations create hydrodynamical perturbations of the little bang --
  the  sounds.  The long-wave ones can survive until the
    freeze-out time   without significant damping  and are observed experimentally,
    in the correlation functions of the secondaries. These observations are  in excellent agreement with the
    hydrodynamics  (see, e.g., Refs.~\cite{Staig:2011wj,Gale:2013da}),
    and this ensures existence of the sound in the near-$T_c$ matter.
   [Shorter-wave sounds, which do not survive until freeze-out,
    were not yet observed, although there have been suggestions  \cite{Basar:2014swa} to use ``magneto-sono-luminescence"
    processes  $phonon+photon \rightarrow photon$ (or $dilepton$) to do so.]


There is, however, an important difference between the hydrodynamics
in the heavy-ion collisions (the ``little bang")
 and the early Universe.
The Reynolds number for QGP at RHIC is estimated \cite{Romatschke:2007eb} to be $\mathrm{Re}_{\mathrm{RHIC}}\sim 48\pi$, with the typical length scale
$R_{\mathrm{Au}}\sim 6\, \mathrm{fm}$, the radius of the gold nucleus. Such a small Reynolds number would not allow
instabilities -- creating the turbulence -- to be  developed. In contrast, for the early Universe, at, e.g., the QCD phase transition,
\begin{align}
\mathrm{Re}_{\mathrm{EU}} = \frac{t_{QCD}\cdot c}{R_{\mathrm{Au}}}\,\mathrm{Re}_{\mathrm{RHIC}} \sim 10^{19}\,,
\end{align}
where we take the cosmological horizon to be a typical length scale (i.e., the big bang fireball is of order of 10 km size).
In this case, the turbulence can be fully developed, while   the viscous forces are mostly
irrelevant.

Thinking of other settings in  nature, with a
very large Reynolds number and
 strong turbulence, one may take, as
 an example, the Sun, or stars, in general. In this case,
  the acoustic waves are  generated by the  convection.
The energy spectrum of the acoustic waves was obtained from various models \cite{Musielak}, and its most prominent feature is the  power spectrum with inverse power of momentum, except for
a flat peak at its smallest values
 $k_{\mathrm{IR}}$.

 The analogy between the early Universe and the Sun  cannot be used in a straightforward way, for several reasons.
 First of all, the Sun is  near stationary, with a well-defined source and sink. Second,
 the Sun's plasma is strongly influenced by long-range magnetic fields, forming
 flux tubes   described by magneto-hydrodynamics (MHD).
The QGP near $T_c$
can be described as a plasma with both electric and magnetic objects \cite{Liao:2006ry, Chernodub:2006gu}.
However, the screening length of both electric and magnetic fields is generally  close to the microscopic scale $1/T$.
 Dynamics
of the electric flux tubes do exist, near and below $T_c$, and it can lead to ``string balls" \cite{Kalaydzhyan:2014tfa}.
While those excitations can lead to interesting phenomena, perhaps to sound generation, they clearly cannot be long range, i.e.,
important at distance scales much larger than the microscale $1/T$.

   Finally, let us also mention papers by
  Kovtun\textit{ et al.}  \cite{Kovtun:2011np, Kovtun:2012rj}
and subsequent works, which initiated our interest in sound interactions. A particular  effect
  calculated in these works is the correction to the viscosity due to sounds, i.e., the ``loop viscosity", appearing technically as a sound loop in
  the energy-momentum correlator $G^{xyxy}(k_\alpha)$.  This effect leads us to think about the sound decay and/or GW formation
(although   their  kinematics is different from what we have considered).

We start with an introductory discussion of the main cosmological parameters of both transitions,
the expected frequencies of  gravity waves and methods for their potential observations.
Section \ref{seq_preliminary} contains a preliminary discussion of thermal radiation,
identifying enhancement parameters, and we conclude that GW thermal radiation
is unobservable. In Sec.~\ref{sec_turb}, we introduce the inverse acoustic turbulent cascade
and then discuss the three-wave or decay dynamics.
(Experts in the corresponding subjects can omit those sections.)
The essential new material starts in Sec.~\ref{sec_4wave}, where we turn to a four-wave
kinetic equation, which leads to the inverse cascade. We then consider possible stationary regimes of strong
turbulence in section \ref{sec_strong}, proceeding to time-dependent behavior in Sec.~\ref{sec_Hindmarsh}.
In Sec.~\ref{soundsound_toGW} we turn to the GW generation rate, and we conclude in Sec.~\ref{summary}.

\section{Frequencies, observational methods  and experimental limits on the cosmic gravity waves}

   Let us briefly mention the numbers related to the QCD and EW transitions. Step one is to evaluate
  redshifts of the transitions, which can be done by comparing the transition temperatures $T_{QCD}=170\, \MeV$ and $T_{EW}\sim 100\, \GeV$ with the temperature of the cosmic microwave background $T_{CMB}=2.73\, \mathrm{K}$. This leads to
  \begin{align}
  z_{QCD}= 7.6 \times10^{11},\quad z_{EW} \sim 4 \times10^{14}\,.
  \end{align}
   At the radiation-dominated era, to which both QCD and EW era belong, the solution to Friedmann equations leads to a well-known relation between the time and the temperature \footnote{Note that we use not gravitational but particle physics units, in which c=1 but the Newton constant  $G_N=1/M_p^2$.},
   \begin{align}
   t=\left({ 90 \over 32\pi^3 N_{DOF}(t)}  \right)^{1/2} {M_P \over T^2}\,,     \label{eqn_t}
   \end{align}
  where $M_P$ is the Planck mass and $N_{DOF}(t)$ is the effective number of bosonic degrees of freedom (see details in, e.g., Particle Data Group big bang cosmology).

   Plugging in the corresponding $T$, one finds the time of the QCD phase transition to be $t_{QCD}=4\times 10^{-5}\,s$ and electroweak $t_{EW}\sim 10^{-11}\,s$.
   Multiplying those times by the respective redshift factors, one finds that the $t_{QCD}$ scale today corresponds to
   about $3\times 10^7\,s=1 $  year, and the electroweak to $5\times 10^4\,s = 15$ hours.

    The cosmological horizon provides a natural infrared cutoff on the gravitational radiation wavelength. At the radiation-dominated era, it is inversely proportional to time, so the estimates above give a cutoff on the periods of the gravitational waves in the present time.
    GWs from the electroweak era are expected to be searched for by future space GW observatories such as eLISA:
 discussion of their potential sensitivity can be found elsewhere. The observational tools for the GW at the period scale of $years$ are based on the long-term
monitoring of the millisecond pulsar phases, with subsequent correlation between all of them. The basic idea is that
when the GW is falling on Earth and, say, stretches distances in a certain direction, then in the orthogonal direction, one expects
distances to be contracted. The binary correlation function for the pulsar time delay is an expected function of the angle $\theta$ between them on the sky. There are existing collaborations -- North American Nanohertz Observatory for Gravitational Radiation, European Pulsar Timing Array (EPTA), and Parkes Pulsar Timing Array -- which actively pursue both searching for new millisecond pulsars and collecting the timing data for some known pulsars. It is believed that about 200 known millisecond pulsars constitute only about 1\% percent of their total number in our Galaxy.
We also note that the current bound on the GW energy density for the frequencies of interest,
$f \thickapprox \mathrm{year}^{-1}$, is \cite{Shannon:2013wma}
\begin{align}
\Omega_{\mathrm{GW}} (f = 2.8 \mathrm{nHz})\cdot (h_0/0.73)^2< 1.3\times 10^{-9} \,, 
\end{align}
where $\Omega_{\mathrm{GW}}$ is, as usual, the total energy density of the GW relative to the critical energy density and
\begin{align}
\Omega_{\mathrm{GW}}(f) = d \Omega_{\mathrm{GW}} / d (\ln f) \,.
\end{align}

This bound should constrain possible models of the GW production in the early Universe. [Note that at the time of the QCD (EW) transition, $\Omega_{\mathrm{rad}}$ is about 4 (15) orders of magnitude larger due to its dependence on the scaling factor $a(t)$, so the aforementioned limit is weaker for those times.]

Rapid progress in the field, including better pulsar timing and  formation of a global collaboration of observers, is expected
to improve the sensitivity of the method, perhaps making it possible in a few-year time scale to  detect
GW radiation, either from the QCD
 big bang GW radiation
we discuss or from colliding supermassive black holes.

\section{Preliminary discussion of sound-to-GW transition} \label{seq_preliminary}
For comparison, let us start with the little bang -- heavy-ion collision. As one of us suggested many years ago \cite{Shuryak:1978ij},
production of penetrating probes -- photons and dileptons -- not only provide a look inside the quark-gluon plasma, but is even somewhat enhanced.
The rate of, e.g., photon production due to the strong Compton scattering and annihilation $qg\rightarrow q\gamma, \bar{q}g\rightarrow \bar{q}\gamma, \bar{q}q
\rightarrow g\gamma$  is
\begin{align}
dN_\gamma/d^4x \sim \alpha \alpha_s T^4
\end{align}
and thus the photon accumulated density normalized to the entropy density of matter $s_{QGP}  \sim T^3$ is of the order of
\begin{align}
{\int dt dN_\gamma/d^4x \over s_{QGP} }\sim \alpha \alpha_s  (t_{life}\, T)\,,
\end{align}
where $t_{life}$ is the fireball lifetime. The small QED and QCD coupling constants in front are thus
partly compensated by large  $(t_{life}\,T) \gg 1$, called the ``macro-to-micro ratio'', which will repeatedly appear below. This factor represents a long accumulation time of the photon production, and
it is about 1 order of magnitude in heavy-ion collisions.

Similar logic holds for the gravitational radiation from matter constituents.
The characteristic microscale of the plasma is its temperature $T$ . At the thermal (the high-frequency)
end of the spectrum, $\omega\sim T$, one finds the fraction of  GW radiation
to the total energy density $T^{00}\sim N_{DOF} T^4$
 to be given by a similar expression,
 \begin{align}
 \Omega_{\mathrm{GW}} \sim \left({T\over M_P}\right)^2 (t_{life}\, T)\,,
 \end{align}
 where the first factor is the corresponding effective gravitational coupling, which
is very small  since $T/M_P\sim 10^{-20}-10^{-17}$ in our case. The macro-to-micro factor is a large enhancement factor, which can be readily obtained from
(\ref {eqn_t}) and in fact contains an inverse of the ratio just mentioned; thus,
\begin{align}
t\,T \sim {M_P \over T}\cdot{1 \over N_{DOF}^{1/2}} \sim 10^{16} - 10^{19}\,. \label{eqn_occ}
\end{align}
This factor cannot, however, cancel all powers of $M_P$ in the coupling factor, so the gravitational radiation directly from plasma particles is strongly suppressed.

While matter is  mostly made of various partons with $k\sim T$, it also contains  long wavelength collective modes,
the hydrodynamical sounds.  Thermal occupations of plasma partons are $n_k=O(1)$, but for sounds,
 even in equilibrium, their occupation factors for small frequencies are much larger, $n_k\sim T/k \gg 1$.

 Out-of-equilibrium phenomena, which we study below, may produce much higher amplitudes of hydrodynamical
 perturbations at small $k$,
in the so-called
 {\em inverse acoustic cascade}. The sound momenta and frequencies are, however, limited from below, and thus the sound intensities $n_k$
 are limited as well.
The most obvious
  infrared  cutoff is by the inverse lifetime of the Universe, $\omega > 1/ t_{life}$: a more precise cutoff is
  due to a collision rate, which we discuss below.

The sound conversion to the GW  happens via a two-to-one transition, and therefore
its rate is enhanced {\em quadratically}, $\sim n_k^2$.
The peak in the sound intensity squared will be repeated in the GW spectrum.
The more it moves to the IR, the stronger the GW signal will be, and the better chances we have
to eventually observe it.

Summarizing this section, only strongly enhanced out-of-equilibrium sounds may potentially produce an observable level of the GW.
The task is  to  estimate  the sound level at the IR end of the dynamical range. To illustrate how
 highly nontrivial it is,  we recall that the loudest sounds on Earth have nothing to do with the
equilibrium conditions but rather with thunderstorms or earthquakes.

\section{Acoustic turbulence} \label{sec_turb}
The idea of turbulence, either driven or free, started from hydrodynamics of fluids. Kolmogorov
proposed the famous stationary power solutions. For the weak turbulence, governed by the
Boltzmann equation, such solutions were developed by Vladimir Zakharov and collaborators, to many different problems,
as summarized in the book \cite{ZLFbook}. 
A turbulent cascade in cosmology was
suggested to appear after the preheating stage of inflation \cite{Micha:2004bv}: for a scalar
field with quartic self-interaction. However, that cascade is  {\em direct}, propagating
into UV, towards the large momenta $k$. Consideration of an {\em inverse} cascade to IR, similar to our
case, was done for scalar theories \cite{Berges:2010ez} as well as recently for
gluons (see, e.g., Ref.~\cite{Berges:2013eia}).
The {\em inverse acoustic cascade} in the strong turbulence regime,
 to our knowledge, was never discussed before.

\subsection{Scenario 1:  Binary decays allowed}

The key features of our theory are nonlinear corrections to the sound dispersion law. We will use notations
\begin{align}
\mathrm{Re}\, \omega_k=c_s k + \delta \omega
\end{align}
and assume that
\begin{align}
\delta\omega= A k^3 + \mathcal{O}(k^5)\,.     \label{dispersion}
\end{align}
The sign of constant $A$ would lead to physically different scenarios due to different sound cascades. Although the coefficient $A$ is not known for the sound near the QCD or EW phase transitions,
it was derived for a strongly coupled plasma of the
$\cal{N}$=4 super-Yang-Mills theory, through the AdS/CFT correspondence. It is widely believed that those should be similar,
at least qualitatively. Without going into details,  the known terms in the sound dispersion curve, up to $\mathcal{O}(k^6)$ accuracy, are \cite{Lublinsky:2009kv}

\begin{align}
 &{\omega\over 2\pi T} =\pm{ \tilde k \over  \sqrt{3}} \left[ 1+ \left({1\over 2}-{\ln 2 \over 3}\right) \tilde k^2-0.088\, \tilde k^4\right]\nonumber\\
&~~~~~~~- {i \tilde k^2\over 3} \left[1-{4-8\ln 2+\ln^2 2 \over 12}\, \tilde k^2 - 0.15\,\tilde k^4\right]\,,\label{SYMdisp}
\end{align}
where $\tilde k \equiv k / (2\pi T)$. The crucial observation is that the $\mathcal{O}(k^2)$ correction in the first bracket of (\ref{SYMdisp}) has a $positive$ coefficient. This allows for three-wave
$1\leftrightarrow 2$ transitions between the sounds -- in particular, a decay of a harder phonon into two softer ones.
Although this is, in principle, known, for completeness let us remind the kinematics of this process.

The momentum conservation  $\vec{k}=\vec k_1+  \vec k_2$ allows us to introduce a parameter $x \in [0, 1]$ and a vector $\vec q_\perp$ such that $\vec k_1, \vec k_2$ will have
 longitudinal components along $\vec{k}$ denoted  by $\vec k_1^\parallel=\vec k\cdot x$, $\vec k_2^\parallel= \vec k\cdot(1-x)$  and
the transverse ones $\vec k_{1,2}^\perp=\pm \vec q_\perp$, where plus (minus) are for $\vec k_1$ ($\vec k_2$).
The energy conservation,
\begin{align}
\omega(k)=\omega(k_1)+\omega(k_2)\,,
\end{align}
can be  simplified using the fact that the dispersive correction is small in the range which we are interested,
\begin{align}
\sqrt{A} k \ll 1\,.
\end{align}
  Realizing
that the transverse momentum is proportional to this, and thus that it is also small,  one may simplify energy conservation
further.  The resulting value of the transverse momentum, for a given value of longitudinal momentum fraction $x$, is
\begin{align}
 {q_\perp \over k} = (\sqrt{A} k)  \sqrt{6}x (1-x)\,.
 \end{align}
One can further argue that, due to the Goldstone nature of sounds, their interaction matrix element at small momenta (IR) must be proportional to the product of all momenta,
\begin{align}
\mid V(k,k_1,k_2)\mid_{{}_{\mathrm{IR}}}^2=b\cdot  k\cdot k_1 \cdot k_2\,, \label{triple}
\end{align}
where $b$ is a constant. Dynamical and even dimensional arguments \cite{ZLFbook} confirm this result.

Having in mind this matrix element, the phase space of the decay, 
one can write down a kinetic equation including all $1\leftrightarrow 2$ transitions.
The details can be found  in  Ref.~\cite{ZLFbook}.
Let us present here only the final form of the Boltzmann equation with the
assumption of the isotropy of spectra and the angle integrations performed,
\begin{align}
 & { 1 \over 4\pi b} {\partial n_k \over \partial t}= \label{kinetic_KZ}\\
 &\int_0^k dk_1 k_1^2(k-k_1)^2 [ n_{k_1}n_{k-k_1} -n_k (n_{k_1}+n_{k-k_1})] \nonumber\\
 &-2\int_k^\infty dk_1 k_1^2(k-k_1)^2 [ n_{k}n_{k_1-k} -n_{k_1} (n_{k}+n_{k_1-k})]\,. \nonumber
\end{align}
   In spite of a relatively complicated form of the equation, it has simple stationary
  power solutions,  generally known as Zakharov's spectra \cite{ZLFbook}, 
\begin{align}
n_k\sim k^{-s}, \qquad s_{decay}=9/2\,. \label{index_decay}
\end{align}
This power solution is in fact a stable  ``attractor" solution.
Numerical simulations, starting from a variety of out-of-equilibrium distributions, have been shown to approach this
spectrum rather rapidly (again, see Ref.~\cite{ZLFbook}).

Unfortunately, the sign of the flux associated with this cascade is such that it develops in
UV direction, making it irrelevant for  problem under consideration.
 Note that the total energy density contained in the sounds,
\begin{align}
\epsilon_{sound}=\int \omega_k n_k 4\pi k^2 dk\,,
\label{eqn_energy}
\end{align}
is convergent at the UV end.

\subsection{Scenario 2: Four-wave interactions} \label{sec_4wave}
Now we discuss an alternative case,
 when the dispersive correction coefficient in (\ref{dispersion})
is negative, $A<0$, and, therefore, the  binary on-shell decays of sound waves are forbidden.  In this case one should consider the second order processes, i.e. the scattering $2\leftrightarrow 2$,  as well as three-body decays $1\rightarrow 3$ and corresponding inverse processes (which are always permitted by the conservation laws).

  For a relativistic scalar theory with triple $\sim g \phi^3$ and quartic  $\sim \lambda \phi^4$ interactions,
 these processes stem either from nonlocal diagrams $O(g^2)$ or from local ones  $O(\lambda)$.
 When only the latter are 
 present, derivation of the kinetic equation for weak turbulence
 is very straightforward (see, e.g., Ref.~\cite{Micha:2004bv}).
 Yet the former diagrams, $O(g^2)$, when present, are dominant, since $t$-channel exchanges lead to the
 small-angle and large impact parameter collisions with large
 cross sections. This is known for gluons and is also the case for sound waves.

   The four-wave scattering amplitude, the Boltzmann equation itself and its stationary solution are more complicated, and we will not
   repeat here the material covered in the Ref.~\cite{ZLFbook}. Let us only briefly mention the ideas essential for the understanding of the weak turbulence.  The $2\leftrightarrow 2$ scattering amplitude is, schematically, a sum of the type
    \begin{align}
   \sum\limits_{i, j, l, m}     {V^*(k_i\pm k_j,k_i,k_j)V(k_l\pm k_m,k_l,k_m) \over \omega(k_i) \pm \omega(k_j) - \omega(k_i \pm k_j)} \label{treeamplitude}
    \end{align}
    where $i, j, l, m = 1, \ldots, 4$ are four participating particles. 
    For small angles $\theta_i$ relative to the momentum $k$ (the external argument of Boltzmann equation), the denominators
    are
    \begin{align}
    & \omega(k) \pm \omega(k_j) - \omega(k \pm k_j) \approx \nonumber \\
    & c_s k {k_j \over 2 |k \pm k_j|} \theta_j^2 +\delta\omega(k)
    \pm  \delta\omega(k_j) - \delta\omega(k \pm k_j)\,.\label{denominator}
    \end{align}
The scattering amplitude is substituted into the collision integral of the Boltzmann equation, which is then solved by means of the scaling analysis. The difficulty is that the first term in (\ref{denominator}) scales as the first power of momentum, while the energy corrections have a different scaling index,
\begin{align}
\delta \omega (\Lambda k)=\Lambda^\beta \delta \omega (k)\,,
\end{align}
which we assume is $\beta=3$. The issue was resolved by Katz and Kontorovich, who suggested complementing the scaling transformation
of momenta by an additional rotation, such that the angles are rescaled by
\begin{align}
\theta'=\Lambda^{(\beta-1)/2} \theta\,.
\end{align}
  Now all terms in the denominators above have the same index $\beta$.
  This transformation keeps (parts of) the collision integral
  invariant and ultimately leads to an isotropic stationary Kolmogorov-like power solution.  For
  the inverse (particle flow) cascade, we are interested in
  the index $s$ of the momentum density $n_k \sim k^{-s}$, which satisfies the constant flux equation,
  \begin{align}
  & -3s + 4m - 3\beta -1 - (\beta+1)\cdot{d-1\over 2}~~~~~~~~~~~~~~~~~~\nonumber\\
  & ~~~~~~~~~~~~~~~~~~~~~~+3 (\beta-1)\cdot{d-1\over 2} +4d =0\,. \label{eqn_index}
  \end{align}
 Here the index  $m$ is the index of the triple vertex, $m=3/2$.
  The first two terms are obvious -- there are three densities and four triple vertices (since we take a square of the amplitude); the third one comes from the energies in the denominator of (\ref{denominator}) and the energy conservation condition, the fourth (fifth) comes from the longitudinal (transverse) momentum conservation condition, and others have to do with the phase space
  integration measure. Note that one should take special care of the argument of the energy conservation under Katz-Kontorovich
  transformation and angular integrations, which produce the last two $\beta$ terms.
  Substituting the space dimension $d=3$ and the index $\beta=3$ of $\delta \omega$, one gets
  \begin{align}
s_{nondecay}=10/3\,. \label{index_nodecay}
\end{align}
     (Another power solution   of the  Boltzmann equation -- the energy flux solution -- has an opposite sign of the flow,
  to UV, which we thus disregard.)

Since the obtained index is in the segment $3< s < 4$, the energy integral (\ref{eqn_energy}) is dominated by the UV end and is thus irrelevant, while
the particle number
\begin{align}
 N= \int  n_k 4\pi k^2 dk \label{eqn_N}
 \end{align}
 $is$ dominated by the IR end.
  Such cascades, driven by
particle number normalizations, are usually called the ``particle number cascades".

\subsection{Scenario 2: Strong turbulence} \label{sec_strong}

This is not the end of the story because  growing particle density at small $k$  eventually
violates the applicability condition of weak turbulence, $n_k\ll 1/\lambda$. So, at the IR end, the
physics is in the regime of strong turbulence, in which consideration of higher order diagrams is required.
To our knowledge, this question was never considered 
in the case of sounds.

The strong turbulence regime was studied in the case of relativistic $\lambda \phi^4$ theory by Berges and collaborators
\cite{Berges:2010ez, Berges:2008wm}, who derived a renormalized inverse cascade, with modified indices.
Importantly, those were  confirmed by direct simulations, in $d=3$ and $4$ spatial dimensions \cite{Berges:2010ez}.

\begin{figure*}[t]
  \begin{center}
\subfigure[\label{diagrams::a}]{\includegraphics[width=6cm]{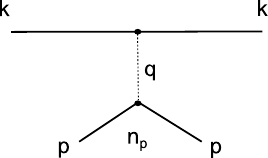}}\hspace{2.5cm}
\subfigure[\label{diagrams::b}]{\includegraphics[width=5cm]{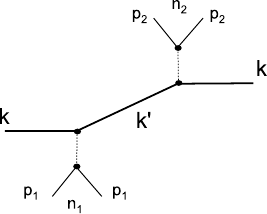}}
   \caption{Forward scattering diagrams corresponding to the (a) quartic and (b) sextic terms in the Hamiltonian (\ref{eqn_Hc}).}
  \label{diagrams}
  \end{center}
\end{figure*}

 The core of their theory is that
the rescattering diagrams can be included in a rather elegant way, via
a renormalized effective self-interaction coupling,
\begin{align}
\lambda_{eff}^2= { \lambda^2 \over ( 1+\Pi(\lambda, k))^2}\,.
\end{align}
At small $k$, $\Pi \gg 1$, so we can neglect 1 in the expression above. Therefore, its scaling index $\Delta$, defined by
\begin{align}
\Pi(k) = \xi^\Delta \Pi(\xi k, \xi \omega_k)
\end{align}
enters the Boltzmann equation, the  expression for  the particle flux and the final equation for the index.
For the $d=3$ case, it is  simply
\begin{align}
\Delta=s\,, \label{eqn_delta}
\end{align}
i.e., the index of the density.  (Density  appears linearly in $\Pi$; other factors cancel.)
Omitting details, the equation for the index then reads
\begin{align}
-4-2\Delta + 3s=0\,.
\end{align}
In the weak turbulence regime, $\Pi\ll 1$, and one should exclude $\Delta$. The index then is $s_{weak}= 4/3$.
However, in the opposite strong turbulence case, one should use (\ref{eqn_delta}), and the index is renormalized to
another -- much larger -- value
\begin{align}
s_{strong} = 4 \qquad \mathrm{(scalar)}\,.
\end{align}
This was the value which was indeed observed in numerical simulations \cite{Berges:2010ez}.

The case of gluon cascade offers some further suggestions and intuition. While it also has a triple vertex and
is dominated by the small-angle scattering, the impact parameter in this case is dominated by
the Debye screening length $b^2<1/M_D^2$ produced by scattering of a virtual gluon on
the ambient plasma, and thus depending on the gluon density.

Let us now try to apply the same logic for the acoustic turbulence.
The main physics idea is that due to the particle
forward scattering on others in the medium, it gains an 
additional correction to its energy, which we will denote by $\delta' \omega$ (with a prime, to distinguish it from the original $\delta \omega$).
Its scaling index is then denoted by $\beta'$. In the strong turbulence regime one expects the rescattering effect
to become dominant, $\delta' \omega \gg \delta \omega$, and hence one should replace $\beta$ by $\beta'$
in the index equation.

Classical perturbation theory, as described in, e.g., Chapter 1 of \cite{ZLFbook}, starts from a Hamiltonian
of the type
\begin{align}
H=\omega b b^* + {V\over 2} (b^2 b^*+ b^{*2}b)+ {U \over 6} (b^3+ b^{*3}) + \ldots
\end{align}
including the wave amplitude $b$ (for brevity, we drop momentum indices here and below) and the triple vertices $V$ and $U$.
In case of nondecay, the triple vertices are irrelevant and can be eliminated by
the canonical transformation
\begin{align}
b=c+ {V \over 2\omega} c^2 - {V \over \omega} c c^* - {U \over 6 \omega} c^{*2}+ \mathcal{O}(c^3)\,,
\end{align}
where $c$ are new amplitudes. 
The new Hamiltonian is then rewritten as
\begin{align}
H=\omega c c^*-{1 \over 4}{\tilde V^2 \over \omega}c^2 c^{*2} + {\bar V^4 \over \omega^3} c (cc^*)^2 c^* + \mathcal{O}(c^7)\,,
\label{eqn_Hc}
\end{align}
where $\tilde V^2 \equiv V^2 + 5 U^2 / 9$  and $\bar V^4 \equiv (2 V^2 U^2 - 3 U V^3 - 27 V^4)/18$.
The next step is to use statistical description, eliminating rapidly varying terms and leaving only slowly changing
correlation functions such as $\langle c_k c^*_{k'}\rangle =n_k \delta(\vec k -  \vec k')$. The second quartic term in (\ref{eqn_Hc})
gives the $2\rightarrow 2$ scattering amplitude; its square appears in the corresponding kinetic equation.

For a generic triple vertex $\tilde V$,  this second term also gives rise to the forward scattering amplitude, Fig.~\ref{diagrams::a}, which can be
reinterpreted as a perturbative correction to the wave energy due to the particle scattering on all others,
\begin{align}
\delta' \omega \sim \int_p{\tilde V^2 \over \omega} n_p d p
\end{align}
(in the spirit of an effective potential for
slow neutrons in ordinary or nuclear matter).
The kinematics of the forward scattering makes two momenta, contributing to the vertex being identical and thus the remaining one being zero. So, naively, if one of the momenta in $V_{k p q}\sim \sqrt{k\cdot p\cdot q}$ vanishes, then the amplitude of the process is zero. However, the denominator in (\ref{treeamplitude}) also vanishes and, applying the l'Hospital's rule with $q\to 0$, one can show that the total expression (the amplitude) is finite. We do not evaluate the absolute magnitude of $\delta' \omega$, only its scaling index,
 \begin{align}
 \beta' =2m-s -1 + 3 =5- s\,.
 \end{align}
Here the $2m$ corresponds to $\tilde V^2$, $s$ to the density $n_p$, and $-1$ to the scaling of the denominator; hence $q$ in (\ref{treeamplitude}), the last term, comes from the integration measure over $\vec p$.  Then we substitute this into the index equation (\ref{eqn_index}) instead of $\beta$ and get a corrected
index for the  strong turbulence
\begin{align}
s_{strong}=4 \label{strong}\,,
\end{align}
corresponding to a flat sound power spectrum.

Here we calculated the index of the diagram, Fig.~\ref{diagrams::a}, and not the diagram itself. In case there is a fine-tuning of the parameters leading to a vanishing contribution of this diagram (which we cannot exclude \textit{a priori}), one should focus on the third term of
 (\ref{eqn_Hc}). It generates a nonzero
 forward scattering and correction to the energy of the order ${\bar V^4 \over \omega^3} n^2 $, from a
 scattering  on $two$ particles [see Fig.~\ref{diagrams::b}].
 The intermediate wave is not collinear with the original one, so in this kinematics $V$ and $U$ do not vanish.
In this case, the index for $\delta' \omega$ will be
\begin{align}
\beta' =4m-2s -4\beta +2(2+\beta) =10-2 s\,,
\end{align}
where, again, the $4m$ corresponds to $\bar V^4$, $2s$ to two densities, and $-4\beta$ to frequencies in the denominator and in the energy conservation condition, and the last term comes from the angular integral.
We substitute it into the index equation (\ref{eqn_index}) instead of $\beta$ and obtain an even larger index
\begin{align}
s_{strong}=6 \qquad \mathrm{(subleading)}.\label{subleading}
\end{align}
At this point, since considering all competing mechanisms and diagrams would go beyond the scope of this paper, we just conjecture that 6 is   the largest possible index.

In summary, we suggest that the strong acoustic turbulence can be considered similarly to the scalar and gluon ones,
with the impact parameters of scattering determined self-consistently, by higher order
rescattering processes. Dedicated theoretical studies and numerical simulations are required in order
to check if the proposed index (\ref{strong}) is correct.  If so, or even if it is different but still, say, large enough, $6\geq s_{strong}\geq 4$,
this would enhance $n_k$ and increase the GW intensity by a huge factor.

  \subsection{Scenario 2: Time evolution}\label{sec_Hindmarsh}
 In the regime where external sources and sinks are switched off, the power Kolmogorov spectra
are represented by self-similar propagating solutions of the type
\begin{align}
n_k = \hat{t}^{-q} f_s[\hat{t}^{-p} \hat{k}]  =  \hat{ t}^{-q} f_s[\xi]\,,          \label{selfsim}
\end{align}
where the $\hat{t}$ and $\hat{k}$ are dimensionless time and momenta, respectively, normalized
to the collision rate at some normalization momentum $k_0$ and $\hat{k}=k/k_0$.
With such normalization, the profile function $f_s[\xi]$ has a maximum at $\xi\sim O(1)$.

For 
the inverse acoustic cascade with four-wave interactions, the indices are
\begin{align}
p=-1,\qquad q=-3\,,
\end{align}
for derivation see chapter 4.3 of \cite{ZLFbook}. The negative sign for the indices means that the
profile $f_s$, defining the sound spectrum, moves toward small $k$ in scale variables $\log(k), \log(t)$
 at later time.

 Note that the integral (\ref{eqn_N})
is conserved for this solution, so it is a kind of ``soliton" made of $N$ interacting sound waves,
propagating in the scale (logarithmic) variables.
 This particle number $N$ is the only information one needs to know from
the early time when the sound was generated.

This self-similar solution is valid  for the weak turbulence regime. As we already discussed, at sufficiently
small $k$, $n_k$ becomes so large that the regime must change to the strong turbulence.
A simple self-similar solution perhaps might not be enough
if the index $s_{strong}\geq 4$, since in this case $both$  integrals $E$ (\ref{eqn_energy}) and $N$ (\ref{eqn_N}) will be dominated by the IR scale:
conservation of both by a $single$  self-similar solution is  not possible, so we
cannot suggest a scenario for the time-dependent solution at this time.
Propagation of sound waves, with all sources and sinks switched off, in a strong turbulence regime
requires additional studies. If the overpopulation of the IR scale in scalar and gluonic cascades leads to the formation of a condensate, it would also be interesting to study the latest stages of the sound turbulence, which may (hypothetically) evolve into a finite number of very loud long-wave sound waves.

  \begin{figure}[t]
  \begin{center}
  \includegraphics[width=8cm]{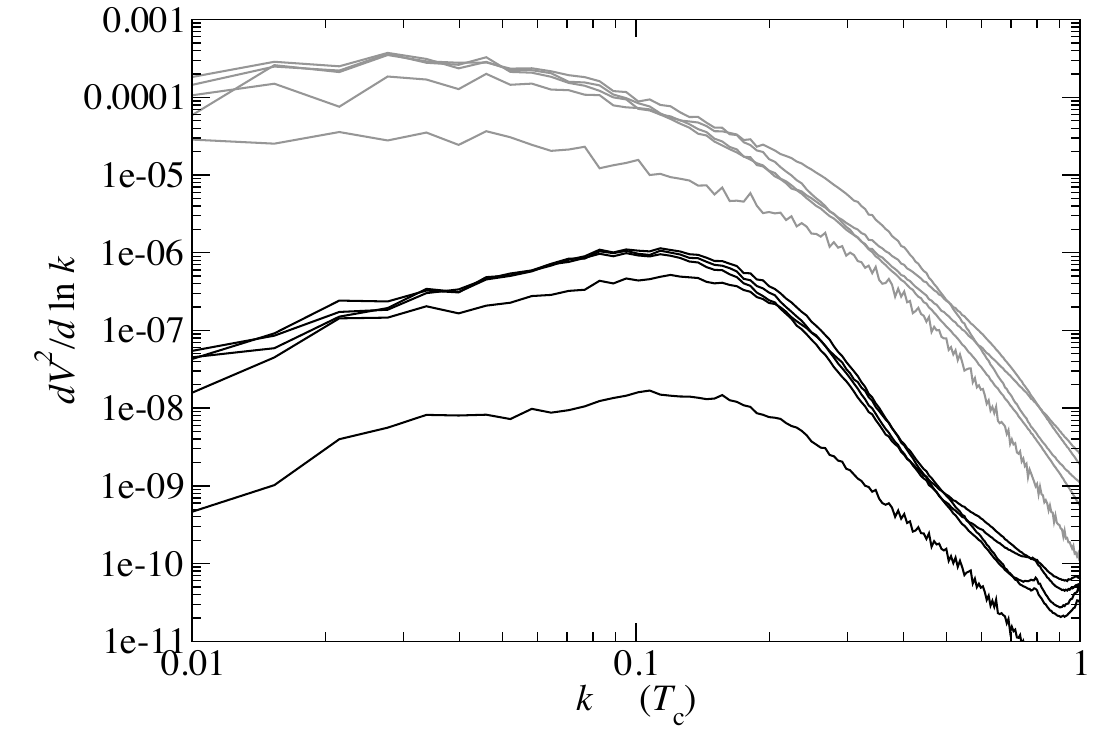}
   \caption{(From Ref.~\cite{Hindmarsh:2013xza}) Power spectrum of the velocity squared versus the
   (log of) the
   wave number $k$. The grey upper curves are for sounds, from bottom to top as time progresses, $t=600, 800, 1000, 1200, 1400\, T_c^{-1}$. The black curves in the bottom are for rotational excitations.}
  \label{fig_hindmarsh}
  \end{center}
\end{figure}

Let us return to the discussion of the initial sound generation, with another look at the results of the numerical simulations
done in Ref.~\cite{Hindmarsh:2013xza}.   Figure~\ref{fig_hindmarsh},
reproduced here from this work, shows
the spectrum of the fluid velocity squared over the log of momentum,
$dV^2/d\log k$.

The first important statement stemming from these spectra
is   that the hydrodynamic perturbations are  dominated by the sound modes (grey curves above), while  the rotational ones (solid curves below) are  suppressed by several orders of magnitude. It is not known how universal this feature is,
but let us accept it for now.

  The spectra in Fig.~\ref{fig_hindmarsh} have
a shallow maximum at $kT\sim 0.03$ corresponding to a
characteristic dynamical
scale of the simulation, the distance between bubbles. Should this calculation be extended to smaller $k$, we think it is inevitable that
the spectrum will be exponentially cut off in the IR. Spectra at subsequent time
moments show no visible tendency of movement of the maximum. We attribute this to the fact
that the total time of the simulation  is simply not enough time for the sound cascade -- and self-similar solution --
to develop.

Note that the typical magnitude of $v^2$ in this simulation is $10^{-4}$ (in relativistic units, with the speed of light $c=1$).
Results of these simulations provide, in principle, the initial sound power spectrum, 
from which the inverse acoustic cascade may start evolving.
Since we expect it to start as weak turbulence in a self-similar form  (\ref{selfsim}),
we only need to know the conserved $N$.
The energy of the sound waves, to the second order, is the unperturbed
density of matter times the fluid velocity squared $(\epsilon+p)_0 V^2$. So one can relate
this spectrum to the sound wave occupation numbers via
\begin{align}
(\epsilon+p)_0 {dv^2  \over d \log k}  \sim 4\pi \omega_k n_k  k^3 \,.
\end{align}
The approximately flat observed left-hand side shows that the effective initial value of the index is close to 4
(of course, only in a limited range of scales and time). Then it is supposed to become the weak turbulence,
and the slope for the curve would be $s_{weak}-4= -2/3$, while the left end of the curve, in the lower $k$ region,
 enters the strong turbulence regime with the slope $s_{strong}-4 = 0 $, i.e., stays flat.
 If $s_{strong}-4>0$, or even 2 as we included as a possibility, the energy spectrum will start growing
 toward small $k$.

\begin{figure*}[t]
  \begin{center}
\subfigure[\label{fig_sketch::a}]{\includegraphics[width=5cm]{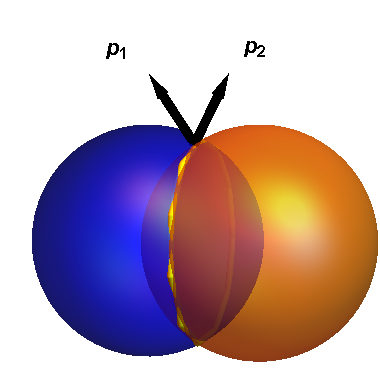}}\hspace{2.5cm}
\subfigure[\label{fig_sketch::b}]{\includegraphics[width=7cm]{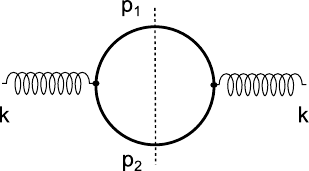}}
   \caption{(a) Sketch of the collision of two sound waves (b) The diagram and the cut described in the text. External legs are gravity waves (gravitons), and the sounds (phonons) are in the loop.}
  \label{fig_sketch}
  \end{center}
\end{figure*}


\section{Generation of gravity waves} \label{soundsound_toGW}
\subsection{The spectral density of the stress tensor correlator}
General expressions for  the GW production rate are well known, and we will
not reproduce them here, proceeding directly to the main object,
the two-point {\em correlator of the stress tensors},
\begin{align}
G^{\mu\nu\mu'\nu'}=\int d^4 x\,  d^4y\, e^{i k_\alpha(x^\alpha-y^\alpha)} \langle T^{\mu\nu}(x) T^{\mu'\nu'}(y) \rangle\,.
\end{align}
Note that while the big bang is homogeneous in space, the 3-momentum can be well defined and conserved, but it is  time dependent.  We will, however, still
treat it as quasistatic, with well-defined frequencies of perturbations, with a cutoff at the lowest end,
$\omega< 1/t_{life}$.

Using hydrodynamical expression for the stress tensor,
\begin{align}
 T^{\mu\nu} =(\epsilon+p)\, u^\mu u^\nu + g^{\mu\nu} p\,,
 \end{align}
and expanding it in powers of a small parameter -- the sound amplitude --
one can identify terms related to the sound wave.
Associating the zeroth order terms with the matter rest frame, one introduces the first order velocities by
\begin{align}
u^\mu=(1,0,0,0)+\delta u^\mu_{(1)}
\end{align}
and one expands the stress tensor to the second order as
\begin{align}
\delta T^{\mu\nu}_{(2)}= (\epsilon+p)_{(0)} \delta u^\mu_{(1)}  \delta u^\nu_{(1)}+ (\epsilon+p)_{(2)} \delta^{\mu 0} \delta^{\nu 0}
    +p_{(2)} g^{\mu\nu}\,.
\end{align}
The correlator is to be coupled to the metric perturbations $h_{\mu\nu}h_{\mu'\nu'}$  and we are interested in
indices corresponding to two polarizations of the GW transverse to its momentum $k_\alpha$.
Such components are only provided by the term with velocities, and thus we focus on
\begin{align}
\int d^4 x\,  d^4y\, e^{i k_\alpha(x^\alpha-y^\alpha)}  \langle \delta u^\mu (x) \delta u^\nu (x)  \delta u^{\mu'} (y)  \delta u^{\nu'}(y)\rangle\,,
\end{align}
where we dropped the overall factor $ (\epsilon+p)_{(0)}^2$ and subscripts ``$(1)$'' for the first order terms.

The next step is to split four velocities into two pairs, for which we use the ``sound propagators",
\begin{align}
 \Delta^{mn} (p^0,\vec{p}) =\int d^4x\, e^{i p_\mu x^\mu} \langle  \delta u^m (x)  \delta u^{n} (0) \rangle\,,
 \end{align}
where we changed indices to the Latin ones, emphasizing that those are only spatial. In these terms, the correlator in question is a loop diagram shown in Fig.~\ref{fig_sketch}(b). Similar loop diagrams were derived and discussed in connection
with fluctuation-induced or loop corrections to hydrodynamical observables: for a recent review of the results,
standard definitions and relations, see \cite{Kovtun:2012rj}.

Time-dependent Green's functions can be chosen differently depending on the assumed boundary conditions on the time dependence.
The most natural Green's functions for the sounds are the retarded one $\Delta_R$, which only has poles in a half of the complex energy $E=p^0$ plane,
corresponding to the sound dissipation, and the symmetric one $\Delta_S$,  which has all four possible poles. In equilibrium, they are related to each other by the so-called Kubo-Martin-Schwinger (KMS) relation ($E=p^0$),
\begin{align}
 -\Delta_S=(1+2 n_B(E)) \Im \Delta_R \underset{E\ll T}{\approx} {2T \over E} \Im \Delta_R\,, \label{KMS}
 \end{align}
where $n_B(E)$ is the equilibrium  Bose distribution.  This expression shows that $\Im \Delta_R$ corresponds to a single phonon quantum, and the $\Delta_S$ to a wave with
proper occupation numbers. It also suggests generalization to an out-of-equilibrium case that we will use, i.e.,
introduction of the new rescaled function
\begin{align}
 -\tilde\Delta_S= 2 n(E)  \Im \Delta_R\,,
 \end{align}
containing  out-of-equilibrium occupation  number $n(E)$, which is assumed to be much larger than the quantum term 1 in (\ref{KMS}),
which is therefore dropped. The explicit expression to be used takes the form
\begin{align}
 \tilde\Delta^{m n}_R= {1 \over (\epsilon+p)_{(0)}} {p^m p^n \over p^2} { E^2 \over
(E^2- p^2 c_s^2)+ i\tilde\gamma p^2 E}\,,\label{propagator}
\end{align}
where notations are three-dimensional, e.g. $p^2=\vec{p}^2$. The dissipation lifetime parameter is related to the shear viscosity
\begin{align}
\tilde\gamma={4 \over 3}\cdot{\eta  \over \epsilon+p}\,.
\end{align}
Now one can perform the Fourier transformation and represent the correlator as a standard field theory loop diagram. The imaginary part
of the correlator, as usual, corresponds to the unitarity cut of the loop into product of two complex conjugated parts, or the probability
of the corresponding sound merging process,
\begin{align}
 &{\Im G^{m m' n n'}(k) \over (\epsilon+p)_{(0)}^2}  = \\
 &\int {d^4p \over (2\pi)^4} n(p^0)\,\Im\tilde\Delta^{mm'}_R(p) n(k^0-p^0)\,\Im\tilde\Delta^{nn'}_R(k-p)\nonumber
\end{align}
 Multiplied by the Newton coupling constant and taken on shell, $k_\alpha^2=0$,
this will give us the rate of the  $sound$+ $sound$ $\rightarrow$ GW process. Note that the unitarity cut also puts both sound lines on shell.

\subsection{Sounds to GW: Kinematics}

One sound wave obviously cannot produce a GW, for the following reasons: (i) The dispersion relation for the sound is $\omega=c_s k$, which is different from that of the GW, $\omega=k$; (ii) polarization of the sound wave is a longitudinal vector, while it should be a transverse tensor for the GW.

Two on-shell sound waves can accomplish this. Using notations
$ p_1^\mu+p_2^\mu=k^\mu$, one writes the GW on-shell condition $(k^\mu)^2=0$ as
\begin{align}
c_s^2(p_1+p_2)^2=p_1^2+ p_2^2+2p_1p_2 \cos(\theta_{12})\,,
\end{align}
where $c_s$,$\theta_{12}$ are the sound velocity and an angle between the two sound waves, respectively.
In terms of such an angle, there are two extreme configurations. The first is a ``symmetric case",  $p_1=p_2$,
corresponding to a minimal angle. 
For $c_s^2=1/3$, this angle is $\theta_{12} =109\degree$. The second, the ``asymmetric case", corresponds to anticollinear vectors $\vec p_1,\vec p_2$,  $\theta_{12} =180\degree$. An important difference from the usual textbook relativistic-invariant cases is that various $\theta_{12}$ are allowed by kinematics in our case, not only $\theta_{12} =0\degree$, which is due to the fact that $c_s < 1$.

 Since the sources of sounds are of microscopic size, $\sim 1/T$, much smaller than the time $t$ of observations,
their sound waves have the form of spherical pulses expanding with the speed of sound.
A sketch of the intersection of two such sound spheres is shown in Fig.~\ref{fig_sketch}: it is clear that
the angle between the sound momenta runs with time over the region allowed for the GW formation.

However, at least at the momentum range in which sounds are weak and the lowest order process $2\rightarrow 1$
dominates the GW production, one may not think about specific hydrodynamical configurations
but simply view it as an incoherent set of plane waves with certain occupation number $n_k$.

\subsection{GW generation rate}
We proceed to the calculation of the ``unitarity cut" of the stress tensor correlator, in which both
sound propagators are taken on shell,
\begin{align}
E_p=\pm c_s p -{i \over 2} \tilde\gamma p^2 \label{on_shell}\,.
\end{align}
One can check that the viscous damping is small, $\tilde\gamma k\ll 1$, so it only needs to go around a pole on the real axis
 in the correct way.
The matrix element is given by a sum over the GW polarizations,
\begin{align}
 \langle \Im G \rangle = \sum\limits_{i=+, \times} \epsilon^{* m n}_{i}\,\Im G_{m m' n n'}\,\epsilon^{m' n'}_{i}\,,
\end{align}
where the polarization matrices can be chosen to be
\begin{align}
\epsilon^{m n}_{+}=\frac{1}{\sqrt{2}}\left(
\begin{array}{c c c c}
0&0&0&0\\
0&1&0&0\\
0&0&-1&0\\
0&0&0&0
\end{array}
\right),
\epsilon^{m n}_{\times}=\frac{1}{\sqrt{2}}\left(
\begin{array}{c c c c}
0&0&0&0\\
0&0&1&0\\
0&1&0&0\\
0&0&0&0
\end{array}
\right)\nonumber
\end{align}
in the transverse traceless gauge, for a plane wave propagating along the third coordinate. Alternatively, one can use a more general standard replacement for the sum,
\begin{align}
 &\sum\limits_{\mathrm{polar.}} \epsilon^*_{m n}\,\epsilon_{m' n'} = \frac{1}{2}\left[\left(\de{m}{m'}\de{n}{n'}+\de{m}{n'}\de{n}{ m'}-\de{m}{ n}\de{m'}{ n'}\right)\right.\nonumber\\
 &~~~~~~~~~~~~~~~~~-\left(\de{m}{ m'}\kh_n \kh_{n'} + \de{m}{ n'}\kh_n\kh_{m'}-\de{m}{ n}\kh_{m'}\kh_{n'}\right)\nonumber\\
 &~~~~~~~~~~~~~~~~~-\left(\de{n}{ n'}\kh_m \kh_{m'} + \de{n}{ m'}\kh_m\kh_{n'}-\de{m'}{n'}\kh_{m}\kh_{n}\right)\nonumber\\
&~~~~~~~~~~~~~~~~~~~~~~~~~~~~~~~~~\left. + \kh_{m}\kh_{n}\kh_{m'}\kh_{n'} \right]\,.\label{crocodile}
\end{align}
Next, the loop momentum integral is customarily rewritten as
$ \int d^4 p_1 d^4p_2 \delta^4(p_1+p_2-k)... $, and the integral over the energies is taken first using the poles of the denominator. The pole residua are the numerator on shell (\ref{on_shell}) divided by the usual $2E_p=2c_s p$,
as for a relativistic particle. Eliminating the integral over $\vec p_2$ and three delta functions one is left
with a single delta function expressing conservation of energy in the process,
\begin{align}
\delta\left[k-c_s p_1 - c_s \sqrt{p_1^2+ k^2-2 p_1 k \cos \alpha_{1k}}\right]\,,\label{delta}
\end{align}
where $\alpha_{1k}$ is an angle between the total (GW) momentum $\vec{k}$ and $\vec p_1$.
So far the steps are similar to a standard calculation of the phase space for particle decays,
in which one can go to the c.m. frame, impose a constraint on momenta from the energy conservation, and
reduce the problem to simple angular integrals.
 Unfortunately, in the problem at hand, we deal with a massless graviton, and we also lack relativistic invariance, which makes this
 procedure useless. Therefore, all three integrals, $d^3p_1=p_1^2 dp_1 d\cos \alpha_{1k} d\phi$, should be done explicitly.

Let us first check the integration limits on $p_1$. From the equations on the energy and momentum conservation, one gets
\begin{align}
 \cos(\alpha_{1k})= \frac{1}{2\,p_1}\left( k - \frac{k}{c_s^2} + 2\, \frac{p_1}{c_s} \right)\,,
\end{align}
and demanding it to be within the range $[-1,1]$, one can constrain the momentum $p_1$ to be between the minimal and maximal values,

\begin{align}
p_1^{max}= \frac{1+c_s}{2\,c_s}k,\qquad p_1^{min}= \frac{1-c_s}{2\,c_s}k\,.\label{p1limits}
\end{align}
Zero of the argument of the delta function (\ref{delta}) falls into this range, so one can simply replace all $p_1$ by this zero.

After summing over two polarizations of the GW and taking into account occupation numbers $n(p)$ for the sounds,
the integral can be written as
  \begin{widetext}
 \begin{align}
 \langle \Im G \rangle&= \int n(p_1) n(k/c_s - p_1) {p_1^2 dp_1 \,d\cos \alpha_{1k}\, d\phi}\cdot \frac{c_s + 1/c_s - 2 \cos \alpha_{1k}}{2(c_s \cos \alpha_{1k} - 1)^2} \cdot \delta\left[ p_1 - \frac{k(c_s^2 - 1)}{2c_s \cos \alpha_{1k} - 1}\right]\nonumber\\
 &~~~~~~~~~~\times {c_s^2 p_1^2 \over  2 c_s p_1}\cdot {c_s^2 (k/c_s - p_1)^2 \over 2(k-c_s p_1)} \cdot \frac{1}{2}\left(1- \cos^2 \alpha_{1k}\right)\left[1 - \left( \frac{k-p_1\cos \alpha_{1k}}{k/c_s-p_1} \right)^2 \right]\,,
 \label{integral}
 \end{align}
 \end{widetext}
where the first line contains the Jacobian for the delta function and the second line comes from the sound propagators
(\ref{propagator}) and the summation formula (\ref{crocodile}).

To make sense of the integral (\ref{integral}), which determines the GW generation rate,
let us consider three simple cases. If the distribution is flat, $n(p) = \mathrm{const}$, then the
integral (\ref{integral}) is proportional to the volume of the phase space,
\begin{align}
 \langle \Im G \rangle_{p^0} \propto \ddd\frac{\pi k^4 (1-c_s^2)^2}{120\,c_s^2}\,.
\end{align}
In the case of thermal equilibrium, $n(p) \propto p^{-1}$, we get a lengthy expression, which can simplified for $c_s=1/\sqrt{3}$,
\begin{align}
 \langle \Im G \rangle_{p^{-1}} \propto \ddd\frac{\pi k^2}{9} \left( \sqrt{3} - 3\, \mathrm{arccoth}\sqrt{3}\right)\,.
\end{align}
Finally, for the strong turbulence cases (\ref{strong}) and (\ref{subleading}), the integral is given by
\begin{align}
 \langle \Im G \rangle_{p^{-4}} &\propto \ddd\frac{4 \pi}{81 k^4} \left( -\sqrt{3} + 5\, \mathrm{arccoth}\sqrt{3}\right)\,,\\
  \langle \Im G \rangle_{p^{-6}} &\propto \ddd\frac{4 \pi}{1215 k^8} \left( 7\sqrt{3} + 55\, \mathrm{arccoth}\sqrt{3}\right)\,,
\end{align}
respectively.

\section{ The QCD phase transition and out-of equilibrium sounds }
\label{sec_qcd_1storder}
In this section, we discuss briefly the status of the debates on the order of the QCD phase transition.
QCD with massless quarks has chiral symmetry, but in the real world, finite quark masses make it only an approximate symmetry.
Therefore, the transition to the broken phase does not need to be a real phase transition.
We know from lattice gauge theory  simulations
that pure gauge SU(3) theory has the first order deconfinement transition. The other extreme --
 QCD with three massless quarks -- also has the first order  transition, now due to the chiral symmetry restoration.
However, for the real QCD, with physical values of $u, d, s$ quark masses, the lattice results indicate, indeed, a smooth crossover-type transition (for current status of the problem see \cite{Bhattacharya:2014ara, Aoki:2006we} and references therein).

  However, the deconfinement is a more subtle story, with the conclusion being much less obvious.
Following the ``dual superconductor" ideas of 't Hooft and Mandelstam from the 1980s,
 the nature of confinement is the Bose-Einstein condensation of certain
magnetically charged objects -- color monopoles. Del Debbio \textit{et al.} proposed an operator inserting a monopole into the vacuum. This operator has a nonzero vacuum expectation in the confined phase, as shown by the direct lattice
simulation \cite{DelDebbio:1995yf}. The behavior of the monopole Bose clusters, which are interchanged along the Matsubara circle -- also
indicates \cite{D'Alessandro:2010xg} that these objects undergo Bose-Einstein condensation at $T<T_c$. Thus, confinement indeed possesses
certain observable ``order parameters".  (Although in the usual ``electric" formulation of the gauge theory
those are nonlocal, they are local in models attempting ``magnetic" formulation.)
Admittedly, two of the lattice works just mentioned are for pure gauge theories which have
phase transitions, not for QCD-like theories with quarks.
The most accurate lattice simulations which focus on
thermodynamical observables do show smoothening of the critical behavior by quark masses, and
for physical QCD, one finds only a cross-over transition so far, without any visible singularity.
(For a long time, this was related to the fact that pure gauge theory is $Z_N$ symmetric, while theories with
fundamental quarks are not. However, discovery of confinement for gauge theories without center symmetry
nullified this argument.)

Thus, there is no clear answer to the question of
whether the deconfinement transition in physical QCD is a phase transition in the strict sense. 
 One  possible resolution may be  a  ``cryptic" transition, in which there is a  singularity in the order parameter,
 which  in thermodynamical observables, is also present but too weak to be seen with current  numerical accuracy.

  Another option for sound and GW generation is that, while there is no first order transition in QCD, and therefore no mixed phase with macroscopically large bubbles,   there may still exist some
  metastable objects in the near-$T_c$ region with a lifetime large enough
   to cause out-of-equilibrium phenomena and sound generation.
      We recently studied dynamics of QCD strings and found  \cite{Kalaydzhyan:2014tfa} that certain nonperturbative objects, so-called ``string balls", can reach rather large mass in
    metastable states, which under a certain slow cooling, can experience rapid collapse, similar to
    the gravitational collapse,
  due to the attractive self-interaction of QCD strings. Such collapse
   can also generate
   inhomogeneous energy distribution, ``overcooling" and subsequent sound generation.

 The freeze-out in the little bang happens very close to the QCD phase transition region.
 Studies of rapidity correlation among secondaries reveal the existence of clustering of secondaries,
perhaps local remnants of the QGP phase.
   The study of this process leads to the suggestion \cite{Shuryak:2013uaa} --  not yet observed --
   that such QGP clusters should implode at $T<T_c$,  in what was called ``mini-bangs".
Such a process may be a very effective mechanism for transferring energy into sounds.

\section{Summary and discussion} \label{summary}
In this paper, we discussed cosmological production of gravity waves from the sound waves, originating in the
big bang phase transitions.
While most of studies focus on the electroweak transition, we
 emphasized the QCD one. Current progress in pulsar timing/correlation technique may help detect cosmological GW even earlier than EW one, for which large GW detectors have to be built in space.

 As a function of momentum scale $k$,
there should be three distinct stages of the process: (i) initial generation of the sound spectrum at the ``UV root" scale $k\sim T$, (ii)  acoustic turbulent cascade, and
(iii) conversion of sounds into the GW. While stage (i) is highly nontrivial and requires further study,  we argue that the intermediate regime (ii) is reasonably well
understood theoretically. 

The possibility of an inverse acoustic cascade is the main new suggestion of this paper. If it happens, the momentum density
of sound $n_k$ becomes self-focused, from large to small momenta $k$. Since the ratio of the UV and IR scales is
as large as 18 orders of magnitude, and the indices (powers of the ratio) can be near 4 or larger,
the enhancements can by huge.

The possibility of having an inverse acoustic cascade depends on the sign of the sound dispersion curve correction (\ref{dispersion}): Only the negative sign is suitable. Currently, for neither the QCD nor the EW plasma do we know this sign.
Thus, we have two cases and perhaps 50\% chances in each: It may happen in one or the other.

If the case with the inverse cascade occurs,
its index will be known in the weak turbulence regime. Furthermore, we expect the self-similar time-dependent
solution to represent time evolution.
Eventually, the inverse acoustic cascade goes into an $n_k$ so large that
the evolution goes into the regime of
{strong} turbulence.  We provide an estimate for the index, imitating renormalization in the scalar theory \cite{Berges:2010ez}. If true, it  suggests a
large  index  (\ref{strong}) and thus potentially very strong enhancement of the sound wave density at small $k$.
It also suggests that a single self-similar time evolution would no longer be possible.
Clearly, dedicated studies are needed.

  Another main result of the paper is the evaluation of the sound-to-GW transition rate. It
   is based on the realization
 that  its rate  can be calculated using the
one-loop sound diagram for the stress-tensor correlator using standard rules.
Furthermore, this loop diagram can be cut by unitarity, putting both sound waves on shell.
The only needed additional ingredient remains the occupancy factors:
The GW yield is proportional to its square at the appropriate momenta.

A mechanism producing sounds is still not understood. Out-of-equilibrium dynamics of QCD and EW
phase transitions remains far from being understood.
  We argued above that certain order parameters jump at $T_c$;
  small-latent-heat deconfinement transition of the first order is still perhaps possible: If so, there would be
 a mixed phase and bubbles, alight with a relatively small contrast in the energy density between the phases.
   So far, it has been assumed in the literature that bubble walls must collide to produce the sounds. However,
 there is another potential mechanism, well known in hydrodynamical literature, namely, the Rayleigh-type collapse of the QGP clusters at $T<T_c$ \cite{Shuryak:2013uaa}. One more possibility we mention is a crossover transition,
 with only microscopic metastable objects -- e.g., the string balls \cite{Kalaydzhyan:2014tfa} -- producing the
 out-of-equilibrium sounds.

{\bf Acknowledgements.} We are grateful to G.~Falkovich for critical comments on the first version of the paper.
This work was supported in part by the U.S. Department of Energy under Contract No. DE-FG-88ER40388.

\end{document}